\documentstyle{article}
\title{\LARGE Equations of 3-th waves hierarchy}
\author{A.~N.~Leznov\thanks{ Universidad Autonoma del Estado de Morelos, CCICAp,Cuernavaca, Mexico}} \date{}

\newcommand{\rig}[2]{\stackrel{#2\rightarrow}{#1}}

\begin{document}
\maketitle

\maketitle

\begin{abstract}

By the method of discrete transformation equations of 3-th wave hierarchy are constructed.  The main difference compare with the systems connected with $A_1$ algebra consists in the fact that in $A_2$ case there are two different systems of equations of the same degree (the maximal derivatives with respect to space coordinate). In the present paper we construct only
system of equations of this hierarhy of first and second degree. At this moment we don't know
the general method (the type of canonical Hamiltonian operators) and our calculations are sufficiently combersome. 

\end{abstract}

\section{Introduction}

The goal of the present paper is in construction the equations of 3-wave hierarchy in explicit form. All system equations of this hierarchy are invariant with respect two mutually commutative discrete transformation of 3-wave problem \cite{I}. In this introduction present the solution of the same problem in the case $A_1$ algebra follow to the paper \cite{DL}.  

We repeat here briefly the most important punkts from \cite{DL}.

The discrete invertible substitution (mapping) defined as
\begin{equation}
\tilde u=T(u,u',...,u^{r})\equiv T(u)\label{1}
\end{equation}
$u$ is $s$ dimensional vector function; $u^r$ its derivatives of corresponding order with 
respect to "space" coordinates.

The property of invertibility means that (\ref{1}) can be resolved and "old" function $u$
may expressed in terms of new one $\tilde u$and its derivativies.

Frechet derivative $T'(u)$ of (\ref{1}) is $s\times s$ matrix operator defined as
\begin{equation}
T'(u)=T_{u}+T_{u'}D+T_{u''}D^2+...\label{FR}
\end{equation}
where $D^m$ is operator of m-times differentiation with respect to space coordinates.

Let us consider equation
\begin{equation}
F_n(T(u))=T'(u)F_n(u)\label{ME}
\end{equation}
where $F_n(u)$ is s-component unknown vector function, each component of which depend on $u$
and its derivatives not more than $n$ order. It is not difficult to understand that evolution type equation
$$
u_t=F_n(u)
$$
is invariant with respect substitution (\ref{1}). 

Two other equations and its solutions are important in what follows
\begin{equation}
T'(u)J(u)(T'(u))^T=J(T(u)),\quad T'(u)H(u)(T'(u))^{-1}=H(T(u))\label{H}
\end{equation}
where $(T'(u))^T=T_{u}^T-DT_{u'}^T+D^2T^T_{u''}D^2+...$ and $J(u),H(u)$ are unknown $s\times s$
matrix operators, the matrix elements of which are polinomial of some finit order with respect
to operator of differenciation (of its positive and negative degrees).

$J^T(u)=-J(u)$ may be connected with the Poisson structure and equation (\ref{H}) means its invariance with respect to discrete transformation $T$.

The second equation (\ref{H}) determine operator $H(u)$, which after application to arbitrary solution of (\ref{ME}) $F(u)$ leads to new solution of the same system
$$
\tilde F(u)=H(u)F(u)
$$
And thus we obtain reccurent procedure to constuct solutions of (\ref{ME}) from few simple ones.
 
If it is possible to find two different $J_1,J_2$ (Hamiltonian operators) then
$$
H(u)=J_2J_1^{-1}
$$
satisfy first equation (\ref{H}).

In \cite{DL} are presented arguments that Hamiltonian operator it is the sence find in a form
\begin{equation}
J(u)=F_n(u)D^{-1}F_n(u)^T+\sum_{i} A_iD^i\label{JJ}
\end{equation}
where $F_n$ some solution of (\ref{ME}) and $A_i$ some $s\times s$ matrices constructed from
$u$ and its derivatves.

But as was noted above in the case of $A_2$ we  were not able to find two corresponding Hamiltonian operators and thus results of this paper necessary consider as primilinary ones.

\section{Equations and discrete transformations of 3 wave problem}

Algebra $A_2$  has the following Cartan matrix and basic commutation relations between two generators of the simple roots $X^{\pm}_{1,2}$ and its Cartan elements $h_{1,2}$
$$
k=\pmatrix{ 2 & -1 \cr
           -1 & 2 \cr},\quad \pmatrix{ [h_1,X^{\pm}_1]=\pm 2X^{\pm}_1 & [h_1,X^{\pm}_2]=\mp X^{\pm}_2 \cr
[h_2,X^{\pm}_1]=\mp X^{\pm}_1 & [h_2,X^{\pm}_2]=\pm 2X^{\pm}_2 \cr}
$$
Arbitrary element of the algebra may be represented as (up to Cartan element)
$$
f=f^+_{1.1}X^+_{\alpha_1+\alpha_2}+f^+_{0.1}2X^+_{\alpha_2}+f^+_{1.9}X^+_
{\alpha_1}+f^-_{1.0}X^-_{\alpha_1}+f^-_{0.1}X^-_{\alpha_2}+f^-_{1.1}X^-_{\alpha_1+\alpha_2} 
$$
$\alpha_{1,2}$ are the indexes of simple roots. $X^+_{\alpha_1+\alpha_2}\equiv [X^+_2,X^+_1]$.

In these notations the system of equations of 3 wave problem looks as
$$
D_{1,0}f^+_{1.0}=f^+_{1.1}f^-_{0.1},\quad D_{1,0}f^-_{1.0}=f^-_{1.1}f^+_{0.1}
$$
\begin{equation}
D_{0,1}f^+_{0.1}=f^+_{1.1}f^-_{1.0},\quad D_{0,1}f^-_{0.1}=f^-_{1.1}f^+_{1.0}\label{DA}
\end{equation}
$$
D_{1,1}f^+_{1.1}=-f^+_{0.1}f^+_{1.0},\quad D_{1,1}f^-_{1.1}=-f^-_{0.1}f^-_{1.0}
$$
where operatores of differentaion are the following ones $D_{1,0}={c_1\over \delta}\frac{\partial}{\partial t}+{d_1\over \delta}\frac{\partial}{\partial x},D_{0,1}={c_2\over \delta}\frac{\partial}{\partial t}+{d_2\over \delta}\frac{\partial}{\partial x},D_{1,1}=D_{1,0}+D_{0,1}, \delta\equiv (c_1d_2-c_2d_1)$,
$t,x$ two independent arguments of 3-wave problem, $c,d$ independent parameters. 
The discrete transformation of this system are the following ones \cite{I}, in what reader can verified by direct not combersom calculation.

\subsubsection{ $T_3$}

The system (\ref{DA}) is invariant with respect to the following transformation $T_3$
$$
\rig{{f^+_{1.1}}}{T_3}={1\over f^-_{1.1}},\quad \rig{{f^+_{1.0}}}{T_3}=-{f^-_{0.1}\over f^-_{1.1}},\quad \rig{{f^+_{0.1}}}{T_3}={f^-_{1.0}\over f^-_{1.1}}
$$
\begin{equation}
\rig{{f^-_{0.1}}}{T_3}=-f^-_{1.1}D_{1,0}{f^-_{0.1}\over f^-_{1.1}},\quad \rig{{f^-_{1.0}}}{T_3}=f^-_{1.1}D_{0.1}{f^-_{1.0}\over f^-_{1.1}}\label{T3} 
\end{equation}
$$
{\rig{f^-_{1.1}}{T_3}\over f^-_{1.1}}=f^+_{1.1}f^-_{1.1}-D_{1,0}D_{0,1} \ln f^-_{1.1}
$$
where $D_{i,j}={(ic_1+jc_2)\over \delta}\frac{\partial}{\partial t}+{(id_1+jd_2)\over \delta}\frac{\partial}{\partial x}$.

\subsubsection{$T_2$}

The system (\ref{DA}) is invariant with respect to the following transformation $T_2$
$$
\rig{f^+_{0.1}}{T_2}
={1\over f^-_{0.1}},\quad \rig{f^-_{1.0}}{T_2}=-{f^-_{1.1}\over f^-_{0.1}},\quad \rig{f^+_{1.1}}{T_2}={f^+_{1.0}\over f^-_{0.1}}
$$
\begin{equation}
\rig{f^+_{1.0}}{T_2}=-f^-_{0.1}D_{1,1}{f^+_{1.0}\over f^-_{0.1}},\quad \rig{f^-_{1.1}}{T_2}=-f^-_{0.1}D_{1,0}{f^-_{1.1}\over f^-_{0.1}}\label{T2} 
\end{equation}
$$
{\rig{f^-_{0.1}}{T_2}\over f^-_{0.1}}=f^+_{0.1}f^-_{0.1}+D_{1,0}D_{1,1} \ln f^-_{0.1}
$$

\subsubsection{$T_1$}

The system (\ref{DA}) is invariant with respect to the following transformation $T_1$
$$
\rig{f^+_{1.0}}{T_1}={1\over f^-_{1.0}},\quad \rig{f^-_{0.1}}{T_1}={f^-_{1.1}\over f^-_{1.0}},\quad \rig{f^+_{1.1}}{T_1}=-{f^+_{0.1}\over f^-_{1.0}}
$$
\begin{equation}
\rig{f^+_{0.1}}{T_1}=f^-_{1.0}D_{1,1}{f^+_{0.1}\over f^-_{1.0}},\quad \rig{f^-_{1.1}}{T_1}=f^-_{1.0}D_{0,1}{f^-_{1.1}\over f^-_{1.0}}\label{T1} 
\end{equation}
$$
{\rig{f^-_{1.0}}{T_1}\over f^-_{1.0}}=f^+_{1.0}f^-_{1.0}+D_{0,1}(D_{1,1} \ln f^-_{1.0})
$$

\subsubsection{General properties of discrete transformations}

Three above transformations are invertable. This means $f$ may be expressed algebraicaly in terms of $\tilde f$. Exept of this $T_3=T_1T_2=T_2T_1$, what means that all discrete transformation are mutually commutative. This in its turn means that arbitrary discrete transformation may be represented in a form $T=T_1^{n_1}T_2^{n_2}$ \cite{I}.
Thus from each given initial solution $W_0\equiv (f^{\pm}_{1.0},f^{\pm}_{1.0},f^{\pm}_{1.1})$ of the system (\ref{DA}) it is possible to obtain a chain of solutions labelled by two
natural numbers $(n_1,n_2)$ the number of applications of the
discrete transformations $(T^1,T^2)$ to it.

The chain of equations which occur with respect to the functions $(f^-_{1.0},f^-_{1.0},f^-_{1.1})$ correspondingly in the case $T_1,T_2,T_3$ discrete transformation are definitely
two-dimensional Toda lattices. Its general solution in the case of two
fixed ends are well-known \cite{LS}. This fact as it was shown in \cite{I},\cite{tok},\cite{nlm}. 
allow to construct the many soliton solutions of the 3-wave problem in most straightforward way.

\section{Extraction of the evolution (time) parameter}

In system (\ref{DA}) and formulae of discrete transformations $T_i$ the time and space coordinates are not identified. Keeping  in mind that it is always possible to add to the right
hand side of (\ref{DA}) the term proportioanal to space derivative, conserving the initial symmetry we devide the time and space coordinates by the condition
$$
D_{1,0}f=f_t+f_x,\quad D_{0,1}f=f_t-f_x,\quad D_{1,1}f=2f_t
$$
and rewrite (\ref{DA}) in equivalent form
$$
\dot {f^+_{1.0}}=-(f^+_{1.0})'+f^+_{1.1}f^-_{0.1},\quad \dot {f^-_{1.0}}=-(f^-_{1.0})'+f^-_{1.1}f^+_{0.1}
$$
\begin{equation}
\dot {f^+_{0.1}}=(f^+_{0.1})'+f^+_{1.1}f^-_{1.0},\quad \dot {f^-_{0.1}}=(f^-_{0.1})'+f^-_{1.1}f^+_{1.0}\label{DAA}
\end{equation}
$$
\dot {f^+_{1.1}}=-{1\over 2}f^+_{0.1}f^+_{1.0},\quad \dot {f^-_{1.1}}=-{1\over 2}f^-_{0.1}f^-_{1.0}
$$
where $\dot f\equiv f_t,f'\equiv f_x$.

The same it is necessary to do in explicit formulae of discrete transformation (see previous section). For example below we present corresponding expression for $T_2$ discrete transformation (explicit formulae for $T_3$case see in AppendixI). In what follows we always use the following order of vector functions
$(f^+_{1.1},f^+_{1.0},f^+_{0.1},f^-_{0.1},f^-_{1.0},f^-_{1.1})$
$$
\rig{f^+_{1.1}}{T_2}={f^+_{1.0}\over f^-_{0.1}},\quad \rig{f^+_{1.0}f^-_{1.0}}{T_2}=2f^+_{1.1}f^-_{1.1}-2({f^-_{1.1}f^+_{1.0}\over f^-_{0.1}})^2-2{f^-_{1.1}f^+_{1.0}\over f^-_{0.1}}((\ln f^+_{1.0})'+(\ln f^-_{0.1})'),
$$
$$
\rig{f^+_{0.1}}{T_2}={1\over f^-_{0.1}},\quad
\rig{f^-_{0.1}f^+_{0.1}}{T_2}=f^+_{0.1}f^-_{0.1}+2f^+_{1.1}f^-_{1.1}-f^+_{1.0}f^-_{1.0}+4(\ln f^-_{0.1})''+4({f^-_{1.1}f^+_{1.0}\over f^-_{0.1}})'-
$$
\begin{equation}
2({f^-_{1.1}f^+_{1.0}\over f^-_{0.1}})^2-2{f^-_{1.1}f^+_{1.0}\over f^-_{0.1}}(\ln f^+_{1.0})'+(\ln f^-_{0.1})'),\quad \rig{f^-_{1.0}}{T_2}=-{f^-_{1.1}\over f^-_{0.1}},
\label{T22} 
\end{equation}
$$
 \rig{f^-_{1.1}f^+_{1.1}}{T_2}=
{1\over 2}f^+_{1.0}f^-_{1.0}+({f^-_{1.1}f^+_{1.0}\over f^-_{0.1}})^2+{f^-_{1.1}f^+_{1.0}\over f^-_{0.1}}((\ln f^+_{1.0})'+(\ln f^-_{0.1})')-({f^-_{1.1}f^+_{1.0}\over f^-_{0.1}})'
$$
The same procedure it is necessary applicate to the equations of $T_1,T_3$ transformation from the previous section(see Appendix II). 

\section{Equations defining the 3-th waves hierarchy}

To obtain these equations it is necessary to differetiate each equations of substitution
on some parameter $p$ and indroduce coresponding notation $(f^{\pm}_{ij})_p=F^{\pm}_{ij}$
and consider $F$ as function nof $f$ and its space derivatives.
Let us rewrite equation (\ref{ME}) in the case of the transformation (\ref{T22}).  For instance three equations of (\ref{T22}) without derivatives look as
\begin{equation}
\rig{F^+_{0.1}}{T_2}=-{1\over (f^-_{0.1})^2}F^-_{0.1},\quad \rig{F^+_{1.1}}{T_2}={F^+_{1.0}\over f^-_{0.1}}-{f^+_{1.0}\over (f^-_{0.1})^2}F^-_{0.1},\quad \rig{F^-_{1.0}}{T_2}=-{F^-_{1.1}\over f^-_{0.1}}+{f^-_{1.1}\over (f^-_{0.1})^2}F^-_{0.1}\label{89}
\end{equation}
All other (3) equations will be presented below after some necessary comments.  

First of all let us notice that right hand side of (\ref{DAA}) is solution of (\ref{ME})
of degree one. But there is other obvious solution of the same degree. Namely $F_1=f'$.
After some not combersome calculations it is possible verified that the zero degree solution
is the following one $F_0=( (c+b)f^+_{1.1},bf^+_{1.0},cf^+_{0.1},-cf^-_{0.1},-bf^-_{1.0},-(c+b)f^-_{1.1}$, where $c,b$ arbitrary numerical parameters. The most simple way to convince in this is to write down corresponding system of equations and check its invariance with respect $T_i$ discrete transformations. Thus in the zero degree
case we have also two independent solutions, which it will be usefull classify as symmetrical
$c=b$ or antisymmetrical one $c=-b$.

Let us seek solution in the form (compare with the case of $A_1$ algebra \cite{DL})
\begin{equation}
F^{\pm}_{1.1}\to \pm f^{\pm}_{1.1}F^{\pm}_{1.1},\quad F^{\pm}_{1.0}\to \pm  f^{\pm}_{1.0}F^{\pm}_{1.0},\quad F^{\pm}_{0.1}\to \pm f^{\pm}_{0.1}F^{\pm}_{0.1}\label{EX}
\end{equation}
Keeping in mind the last substitution we rewrite (\ref{89}) in form (conserving the same notations for unknown functions)
\begin{equation}
\rig{F^+_{0.1}}{T_2}=F^-_{0.1},\quad \rig{F^+_{1.1}}{T_2}=F^+_{1.0}+F^-_{0.1},\quad \rig{F^-_{1.0}}{T_2}=F^-_{1.1}-F^-_{0.1}\label{892}
\end{equation}
The same equations in the case of $T_1$ and $T_3$ substitution look correspondingly as
\begin{equation}
\rig{F^+_{1.0}}{T_1}=F^-_{1.0},\quad \rig{F^+_{1.1}}{T_1}=F^+_{0.1}+F^-_{1.0},\quad \rig{F^-_{0.1}}{T_1}=F^-_{1.1}-F^-_{1.0}\label{891}
\end{equation}
\begin{equation}
\rig{F^+_{1.1}}{T_3}=F^-_{1.1},\quad -\rig{F^+_{1.0}}{T_3}=F^-_{0.1}-F^-_{1.1},\quad \rig{F^-_{0.1}}{T_3}=-F^-_{1.0}+F^-_{1.1}\label{893}
\end{equation}

Now we would like to explain how looks other three equations follows from (\ref{T22}).
They contain 3 different basical terms $f^+_{i.j}f^-_{i.j},{f^-_{1.1}f^+_{1.0}\over f^-_{0.1}}, 
(\ln f^+_{1.0})'+(\ln f^-_{0.1})'$. Differentiation these terms with respect to independent parameter $p$ and taking into acount (\ref{EX}) leads tothe following terms in corresponding equation
$$
f^+_{i.j}f^-_{i.j}\to f^+_{i.j}f^-_{i.j}(F^+_{i.j}-F^-_{i.j}),\quad  
(\ln f^+_{1.0})'+(\ln f^-_{0.1})'\to (F^+_{1.0})'-(F^-_{0.1})',
$$
$$
{f^-_{1.1}f^+_{1.0}\over f^-_{0.1}}\to {f^-_{1.1}f^+_{1.0}\over f^-_{0.1}}(-F^-_{1.1}+F^+_{1.0}+F^-_{0.1}).
$$
The explicit form of these equations will be presented below.

\section{Equations of the first degree}

In spite of fact that these equations were described above we would like to demonstrate on this example the ideas and technique of calculations. Let us seek solution of this problem in a form 
$$
\dot {\ln f^+_{1.1}}={\nu_{1.1}(f^+_{1.1})'+\gamma_{1.1}f^+_{1.0}f^+_{0.1}\over f^+_{1.1}},\quad
\dot {\ln f^+_{1.0}}={\nu_{1.0}(f^+_{1.0})'+\gamma_{1.0}f^-_{0.1}f^+_{1.1}\over f^+_{1.0}}
$$
$$
\dot {\ln f^+_{0.1}}={\nu_{0.1}(f^+_{0.1})'+\gamma_{0.1}f^-_{1.0}f^+_{1.1}\over f^+_{0.1}},
\quad \dot {\ln f^-_{0.1}}={\nu_{0.1}(f^-_{0.1})'+\gamma_{0.1}f^+_{1.0}f^-_{1.1}\over f^-_{0.1}},
$$
\begin{equation}
{}\label{FD}
\end{equation}
$$
\dot {\ln f^-_{1.0}}={\nu_{1.0}(f^-_{1.0})'+\gamma_{1.0}f^+_{0.1}(f^-_{1.1}\over f^-_{1.0}},\quad
\dot {\ln f^-_{1.1}}={\nu_{1.1}(f^-_{1.1})'+\gamma_{1.1}f^-_{1.0}f^-_{0.1}\over f^-_{1.1}}.
$$
The system (\ref{21}) is written under additional condition that equation for $f^-$ components
may be obtained from $f^+$ by corresponding discrete transformation. For instance after $T_3$ (explicit formulae in Appendix II) discrete transformation first equation of (\ref{FD}) looks as
$$ 
\dot {\ln (f^-_{1.1})^{-1}}={\nu_{1.1}((f^-_{1.1})^{-1})'-\gamma_{1.1}f^-_{1.0}f^-_{0.1}(f^-_{1.1})^{-2}\over (f^-_{1.1})^{-1}}
$$
This is exactly last equation from (\ref{FD}). Each equation (\ref{892}),(\ref{893})
leads to following limitation on parameters $\nu,\gamma$ (it is necessary not forget that $F^+$
coincides with the right hand side of (\ref{FD} and $F^-$ is opposite by sighn to it)
\begin{equation}
\nu_{01}-\nu_{10}=2\gamma_{10},\quad \nu_{01}+\nu_{10}=2\nu_{11},\quad -2\gamma_{11}=\gamma_{10}=\gamma_{01}\label{EX}
\end{equation}
We will not check here that all other equations are satisfied but show what choice of parameters
leads to yet known solutions of the first degree.  
In connection with  (\ref{EX}) solution of the first degree depends on two arbitrary parameters
$\nu_{01},\nu_{10}$. Really it depends only on one parameter ${\nu_{01}\over \nu_{10}}$ because
common factor may be included into time variable. Let us choose at first $\nu_{01}+\nu_{10}=0,\nu_{01}=1,\nu_{10}=-1,\nu_{11}=0,\gamma_{10}=\gamma_{01}=1,\gamma_{11}=-{1\over 2}$. This is exactly the case of (\ref{DAA}).The choise $\nu_{01}=\nu_{10}=\nu_{11}=1,\gamma_{10}=\gamma_{01}=\gamma_{11}=0$ leads to trivial solution
$\dot {f}=f'$. 
Under conditions (\ref{EX}) system (\ref{FD}) is Hamiltonian with Hamiltonian 
$$
H=\nu_{11}(f^+_{1.1}(f^-_{1.1})'-(f^+_{1.1})'f^-_{1.1})+{\nu_{10}\over 2}(f^+_{1.0}(f^-_{1.0})'-(f^+_{1.0})'f^-_{1.0})+{\nu_{01}\over 2}(f^+_{0.1}(f^-_{0.1})'-(f^+_{0.1})'f^-_{0.1})+
$$
$$
2\gamma_{11}(f^+_{1.1}f^-_{1.0}f^-_{0.1}-f^+_{0.1}f^+_{1.0}f^-_{1.1})
$$
with non zero Poisson breakets
\begin{equation}
\{f^+_{1.1},f^-_{1.1}\}={1\over 2},\quad \{f^+_{1.0},f^-_{1.0}\}=1,\quad \{f^+_{0.1},f^-_{0.1}\}=1\label{P}
\end{equation}

\section{Equations of the second order}
  

We will try to find system equations of the second order invariant with respect three discrete transformations $T_i$ in a form 
$$
\dot {\ln f^+_{1.1}}={\nu_{1.1}(f^+_{1.1})''+\gamma_{1.1}f^+_{1.0}(f^+_{0.1})'+\delta_{1.1}f^+_{0.1}(f^+_{1.0})'\over f^+_{1.1}}+R_{11},
$$
where $a\equiv (b+c),R_{ij}\equiv (2a_{ij}f^+_{1.1}f^-_{1.1}+b_{ij}f^+_{1.0}f^-_{1.0}+c_{ij}f^+_{0.1}f^-_{0.1})$.
$$
\dot {\ln f^+_{1.0}}={\nu_{1.0}(f^+_{1.0})''+\gamma_{1.0}f^-_{0.1}(f^+_{1.1})'+\delta_{1.0} f^+_{1.1}(f^-_{0.1})'\over f^+_{1.0}}+R_{10}
$$
$$
\dot {\ln f^+_{0.1}}={\nu_{0.1}(f^+_{0.1})''+\gamma_{0.1}f^-_{1.0}(f^+_{1.1})'+\delta_{0.1}f^+_{1.1}(f^-_{1.0})'\over f^+_{0.1}}+R_{01},
$$
\begin{equation}
{}\label{21}
\end{equation}
$$
-\dot {\ln f^-_{0.1}}={\nu_{0.1}(f^-_{0.1})''+\gamma_{0.1}f^+_{1.0}(f^-_{1.1})'+\delta_{0.1}f^-_{1.1}(f^+_{1.0})'\over f^-_{0.1}}+R_{01},
$$
$$
-\dot {\ln f^-_{1.0}}={\nu_{1.0}(f^-_{1.0})''+\gamma_{1.0}f^+_{0.1}(f^-_{1.1})'+\delta_{1.0}f^-_{1.1}(f^+_{0.1})'\over f^-_{1.0}}+R_{10},
$$
$$
-\dot {\ln f^-_{1.1}}={\nu_{1.1}(f^-_{1.1})''+\gamma_{1.1}f^-_{1.0}(f^-_{0.1})'+\delta_{1.1}f^-_{0.1}(f^-_{1.0})')\over f^-_{1.1}}+R_{11}.
$$
The system (\ref{21}) is written under additional condition that equation for $f^-$ components
coinsides by the form with the corresponding equations for $f^+$ components and may be obtained from them by corresponding discrete transformation. Let us perform first equation (\ref{21}) by
$T_3$ transformation. Using (\ref{T3}) we have
$$
\dot {\ln {1\over f^-_{1.1}}}=\nu_{1.1}({1\over f^-_{1.1}})''-\gamma_{1.1}({f^-_{0.1}\over f^-_{1.1}})'{f^-_{1.0}\over f^-_{1.1}})-\delta_{1.1}({f^-_{0.1}\over f^-_{1.1}})({f^-_{1.0}\over f^-_{1.1}})'+R_{11}+
$$
$$
(b_{11}-c_{11})({f^-_{1.0}f^-_{0.1}\over f^-_{1.1}})'+2a_{11}(\ln f^-_{1.1})''.
$$
In writing of the last expression we have used (\ref{RIG3}) from Appendix.
Comparing this expression ( after trivial manipulations) with the last equation  from (\ref{21}) we obtain additionaly equalities
$$
\nu_{11}=a_{11},\quad \gamma_{1.1}+\delta_{1.1}=(b_{11}-c_{11})
$$
The same consideration in connection with $f^{\pm}_{1.0},f^{\pm}_{0.1}$ equations
from (\ref{21}) lead to the equalities
$$
\nu_{10}=2b_{10},\quad \gamma_{1.0}+\delta_{1.0}=2(c_{10}-b_{10}),\quad \nu_{01}=2c_{01},\quad \gamma_{0.1}+\delta_{0.1}=2(c_{01}-b_{01})
$$

\subsection{Equations without derivatives}

In this subsection we would like to find to what further limitations on parameters of the problem under consideration follows from equatios (\ref{891})-(\ref{893}). For a example
consider third equation (\ref{892})
$$
\rig{F^+_{1.0}}{T_3}=-F^-_{0.1}+F^-_{1.1}
$$
Substituting in this expression all necessary values from (\ref{21}) and (\ref{T22}) we come to the equality
$$
2b_{10}{({f^-_{0.1}\over f^-_{1.1}})''\over ({f^-_{0.1}\over f^-_{1.1}}} +\gamma_{10}{f^-_{1.1})'\over f^-_{1.1}f^-_{0.1}}[-2(f^-_{0.1})'-f^-_{1.1}f^+_{1.0}-{1\over 2}{f^-_{1.0}(f^-_{0.1})^2\over f^-_{1.1}}+{f^-_{0.1}(f^-_{1.1})'\over f^-_{1.1}}]-
$$
$$
\delta_{10}{1\over f^-_{0.1}}
[-2(f^-_{0.1})'-f^-_{1.1}f^+_{1.0}-{1\over 2}{f^-_{1.0}(f^-_{0.1})^2\over f^-_{1.1}}+{f^-_{0.1}(f^-_{1.1})'\over f^-_{1.1}}]'+2a_{10}(\ln f^-_{1.1})''+
$$
$$
(b_{10}-c_{10}){f^-_{1.0}f^-_{0.1}\over f^-_{1.1}}+R_{10}=
-{2c_{0.1}(f^-_{0.1})''+\gamma_{0.1}f^+_{1.0}(f^-_{1.1})'+\delta_{0.1}f^-_{1.1}(f^+_{1.0})'\over f^-_{0.1}}-
$$
$$
R_{01}+{a_{1.1}(f^+_{1.1})''+\gamma_{1.1}(f^+_{1.0}(f^+_{0.1})'+\delta_{1.1}f^+_{0.1}(f^+_{1.0})')\over f^+_{1.1}}+R_{11}
$$
After comparision the coefficients under the terms of the same structure we come to the following relations between parameters of the problem
\begin{equation}
H_M=\pmatrix{ a_{11}=-2c_{10} & b_{11}=a_{10} & c_{11}=-3c_{10}-b_{10} \cr
          a_{10}=c_{10}+b_{10} & b_{10}=b_{10} & c_{10}=c_{10} \cr
         a_{01}=-3c_{10}-b_{10} & b_{01}=c_{10} & c_{01}=-b_{10}-4c_{10} \cr}\label{HM}
\end{equation} 
All other parameters may be represented via 2 ones $c_{10},b_{10}$ as follows
$$
\delta_{10}=4c_{10},\quad \gamma_{10}=-2(c_{10}+b_{10}),\quad 2\gamma_{11}=\delta_{10}-\gamma_{10},
$$
\begin{equation}
2\delta_{11}=-\gamma_{10},\quad \delta_{01}=-\delta_{10},\quad \gamma_{01}=\gamma_{10}-\delta_{10}\label{GD}
\end{equation}
Thus all parameters of the system of the second order are expressed lineary by two independent
ones. Indeed only one parameter is independent because common factor may be included into time variable
The matrix $H_M$ (\ref{HM}) is symmetrical one and as consequence of this fact the system (\ref{21}) may be rewritten in terms of Hamiltonian formalism.

\subsection{Conservation laws}

As a direct consequence of (\ref{21}) and (\ref{GD})(!) the following conservation laws take place  
$$
\dot {(2f^+_{1.1}f^-_{1.1}+f^+_{1.0}f^-_{1.0})}=2a_{11}((f^+_{1.1})'f^-_{1.1}-f^+_{1.1}(f^-
_{1.1})')'+2b_{10}((f^+_{1.0})'f^-_{1.0}-f^+_{1.0}(f^-_{1.0})')'+
$$
$$
\gamma_{10}(f^-_{0.1}f^-_{1.0}f^+_{1.1}-f^+_{0.1}f^+_{1.0}f^-_{1.1})'
$$
$$
\dot {(2f^+_{1.1}f^-_{1.1}+f^+_{0.1}f^-_{0.1})}=2a_{11}((f^+_{1.1})'f^-_{1.1}-f^+_{1.1}(f^-_{1.1})')'+2c_{01}((f^+_{0.1})'f^-_{0.1}-f^+_{0.1}(f^-_{0.1})')'+
$$
\begin{equation}
\gamma_{01}(f^-_{0.1}f^-_{1.0}f^+_{1.1}-f^+_{0.1}f^+_{1.0}f^-_{1.1})'\label{INT}
\end{equation}
We pay attention of the reader that the condition that the system posses conservation laws
lead directly to limitations (\ref{GD}).

Now we would like to use conservation laws for cheking equation with derivatives. For this goal
let us construct difference between shifted by discrete transformation $T_i$ and unshifted one
of equations of integral of motions (\ref{INT}). Let us denote such difference by symbol $\Delta_i$      
$$
\dot {\Delta_i(2f^+_{1.1}f^-_{1.1}+f^+_{1.0}f^-_{1.0})})=2a_{11}\Delta_i((f^+_{1.1})'f^-_{1.1}-f^+_{1.1}(f^-_{1.1})')'+
$$
$$
2b_{10}\Delta_i((f^+_{1.0})'f^-_{1.0}-f^+_{1.0}(f^-_{1.0})')'+
\gamma_{10}\Delta_i(f^-_{0.1}f^-_{1.0}f^+_{1.1}-f^+_{0.1}f^+_{1.0}f^-_{1.1})'
$$
In connection with the equations of discrete transformation $see AppendixI$ $\Delta_i R$ is the derivative with respect to space variables and thus space derivatives may be canseled from both sides of the last equality. For instance in the case of $T_3$ discrete transformation the last expression may be rewritten as
$$
\dot {2(\log f^-_{11})'+{f^-_{1.0}f^-_{0.1}\over f^-_{1.1}}}=2a_{11}\Delta_3((f^+_{1.1})'f^-_{1.1}-f^+_{1.1}(f^-_{1.1})')+2b_{10}\Delta_3((f^+_{1.0})'f^-_{1.0}-f^+_{1.0}(f^-_{1.0})')+
$$
\begin{equation}
\gamma_{10}\Delta_3(f^-_{0.1}f^-_{1.0}f^+_{1.1}-f^+_{0.1}f^+_{1.0}f^-_{1.1})'\label{FIN}
\end{equation}

\subsection{Cheking the last equations of discrete substitution}

After differentiation  the left hand side equation (\ref{FIN}), substituting in it time derivatives from (\ref{21}) and substitung in left hand side $\Delta_3$ from formulae
of Appendix II, we obtain the following expression to be checked
$$
-2({a_{1.1}(f^-_{1.1})''+\gamma_{1.1}f^-_{1.0}(f^-_{0.1})'+\delta_{1.1}f^-_{0.1}(f^-_{1.0})')\over f^-_{1.1}})'-2R'_{11}+{f^-_{1.0}f^-_{0.1}\over f^-_{1.1}}\times
$$
$$
[{a_{1.1}(f^-_{1.1})''+\gamma_{1.1}f^-_{1.0}(f^-_{0.1})'+\delta_{1.1}f^-_{0.1}(f^-_{1.0})')\over f^-_{1.1}}-{2b_{10}(f^-_{1.0})''+\gamma_{1.0}f^+_{0.1}(f^-_{1.1})'+\delta_{1.0}f^-_{1.1}(f^+_{0.1})'\over f^-_{1.0}}-
$$
$$
{2c_{0.1}(f^-_{0.1})''+\gamma_{0.1}f^+_{1.0}(f^-_{1.1})'+\delta_{0.1}f^-_{1.1}(f^+_{1.0})'\over f^-_{0.1}}]=-2a_{11}[2(f^+_{1.1}f^-_{1.1})'+({(f^-_{1.1})''\over f^-_{1.1}})' +
$$
$$
{1\over 2}{(f^-_{0.1})'f^-_{0.1}-(f^-_{1.0})'f^-_{0.1})\over f^-_{1.1}}{(f^-_{1.1})'\over f^-_{1.1}} +{1\over 2}{(f^-_{0.1})''f^-_{0.1}-(f^-_{1.0})''f^-_{0.1})\over f^-_{1.1}}+{1\over 2}(f^+_{10}f^-_{10}+f^+_{0.1}f^-_{0.1})'+
$$
$$
(f^+_{1.0}f^-_{1.0}+f^+_{0.1}f^-_{0.1}){(f^-_{1.1})'\over f^-_{1.1}} +{1\over 2}{f^-_{1.0}f^-_{0.1}\over f^-_{1.1}}{(f^-_{1.0}f^-_{0.1})'\over f^-_{1.1}}+2b_{10}{f^-_{1.0}f^-_{0.1}\over f^-_{1.1}}[-2{(f^-_{1.0})''\over f^-_{1.0}}+{(f^-_{1.1})''\over f^-_{1.1}}+{(f^-_{1.0})'f^-_{0.1}\over f^-_{1.1}}+
$$
$$
2{(f^-_{1.0})'(f^-_{0.1})'\over f^-_{1.0}f^-_{0.1}}-{(f^-_{1.1})'\over f^-_{1.1}}({(f^-_{1.0})'\over f^-_{1.0}}+{(f^-_{0.1})'\over f^-_{0.1}})]+
2{(f^-_{1.1})'\over f^-_{1.1}}f^+_{0.1}f^-_{0.1}+
$$
$$(f^+_{0.1})'f^-_{0.1}-f^+_{0.1}(f^-_{0.1})'-f^-_{1.0}(f^+_{1.0})'+(f^-_{1.0})'f^+_{1.0}+
\gamma_{10}{f^-_{1.0}f^-_{0.1}\over f^-_{1.1}}[{(f^-_{1.1})''\over f^-_{1.1}}+4{(f^-_{1.0})'(f^-_{0.1})'\over f^-_{1.0}f^-_{0.1}}-
$$
$$
2({(f^-_{0.1})'\over f^-_{0.1}}+{(f^-_{1.0})'\over f^-_{1.0}}){(f^-_{1.1})'\over f^-_{1.1}}-
{1\over 2}{(f^-_{0.1})'f^-_{0.1}-(f^-_{1.0})'f^-_{0.1})\over f^-_{1.1}}]+
$$
$$
2((f^-_{1.0})'f^+_{1.0}-(f^-_{0.1})'f^+_{0.1})+{(f^-_{1.1})'\over f^-_{1.1}}
(f^-_{0.1}f^+_{0.1}-f^-_{1.0}f^+_{1.0})
$$
To the greatest astonishment of the author the last equation satisfies identicaly  and gives no further limitations on the parametes of the problem. Thus  equations (\ref{21}) together
with limitations on parameters (\ref{HM}) and (\ref{GD}) solve the problem of construction
system equations of the second order invariant with respect to discrete transformation of the three wave problem.

\section{Outlook}

From the physical point of the view the system (\ref{21}) may be interpreted as three interuction nonlinear Schredinger fields with two arbitrary parameters, which it is possible to connect with mass of the particles and constants of their interuction. Author have no knoledges about applications of these equations to physical problems.  

From mathematical point of view in this paper (and in the previous ones) we have gone in the following way from integrable system ad hoc $\to$ discrete transformation $\to$ resolving of them and observation that result is very near to group representation theory and may understood (and explained) in terms of the last. The most interesting question is it possible to find in group representation theory some objects responsible directly for discrete transformation? If such objects will be found then all theory of integrable systems will be possible to explain as some branch of group representation theory.

\section{Appendix I $T_2$ case}

Formulae $T_2$ substitution (\ref{T22}) in the text leads to 
$$
\tilde f^+_{0.1}\tilde f^-_{0.1}=f^+_{0.1}f^-_{0.1}+2f^+_{1.1}f^-_{1.1}-f^+_{1.0}f^-_{1.0}+
4(\ln f^-_{0.1})''+{4(f^-_{1.1})'f^+_{1.0}+2f^-_{1.1}(f^+_{1.0})'\over f^-_{0.1}}-{6(f^-_{0.1})'f^+_{1.0}f^-_{1.1}+2(f^-_{1.1}f^+_{1.0})^2\over (f^-_{0.1})^2}
$$
$$
\tilde f^+_{1.0}\tilde f^-_{1.0}=2f^+_{1.1}f^-_{1.1}-
{2f^-_{1.1}(f^+_{1.0})'\over f^-_{0.1}}-2{(f^-_{0.1})'f^+_{1.0}f^-_{1.1}+(f^-_{1.1}f^+_{1.0})^2\over (f^-_{0.1})^2}
$$
$$
\tilde f^+_{1.1}\tilde f^-_{1.1}={1\over 2}f^+_{1.0}f^-_{1.0}-{(f^-_{1.1})'f^+_{1.0}\over f^-_{0.1}}+{2(f^-_{0.1})'f^+_{1.0}f^-_{1.1}+(f^-_{1.1}f^+_{1.0})^2\over (f^-_{0.1})^2}
$$
Let us use notation from the main text
$$
R=cf^+_{0.1}f^-_{0.1}+bf^+_{1.0}f^-_{1.0}+2af^+_{1.1}f^-_{1.1},\quad a=b+c
$$ 
and under $\rig{R}{T_i}$ understand $R$ after application to it discrete transformation $T_i$ then from calculation above we obtain
\begin{equation}
\rig{R}{T_2}=R+2(c-b)({f^+_{1.0}f^-_{1.1}\over f^-_{0.1}})'+4c(\ln f^-_{0.1})''\label{RIG2}
\end{equation} 
The similar calculations leads to 
\begin{equation}
\rig{R}{T_1}=R+2(c-b)({f^+_{0.1}f^-_{1.1}\over f^-_{1.0}})'+4b(\ln f^-_{1.0})''\label{RIG1}
\end{equation}
\begin{equation}
\rig{R}{T_3}=R+
(b-c)({f^-_{1.0}f^-_{0.1}\over f^-_{1.1}})'+2a(\ln f^-_{1.1})''\label{RIG3}
\end{equation}
 
\section{Appendix II. $T_3$ case}
$$
\rig{f^+_{1.1}}{T_3}={1\over f^-_{1.1}},\quad \rig{f^+_{1.0}}{T_3}=-{f^-_{0.1}\over f^-_{1.1}},\quad \rig{f^+_{0.1}}{T_3}={f^-_{1.0}\over f^-_{1.1}},
$$
$$
\rig{f^-_{0.1}}{T_3}=-2(f^-_{0.1})'-f^-_{1.1}f^+_{1.0}-{1\over 2}{(f^-_{0.1})^2f^-_{1.0}\over f^-_{1.1}}+{(f^-_{1.1})'f^-_{0.1}\over f^-_{1.1}}
$$
$$
\rig{f^-_{1.0}}{T_3}=-2(f^-_{1.0})'+f^-_{1.1}f^+_{0.1}+{1\over 2}{(f^-_{1.0})^2f^-_{0.1}\over f^-_{1.1}}+{(f^-_{1.1})'f^-_{1.0}\over f^-_{1.1}}
$$
$$
\rig{f^-_{1.1}f^+_{1.1}}{T_3}=f^+_{1.1}f^-_{1.1}+(\ln f^-_{1.1})''+{1\over 2}{(f^-_{0.1})'f^-_{0.1}-(f^-_{1.0})'f^-_{0.1})\over f^-_{1.1}}+{1\over 2}(f^+_{1.0}f^-_{1.0}+f^+_{0.1}f^-_{0.1})+{1\over 4}({f^-_{1.0}f^-_{0.1}\over f^-_{1.1}})^2
$$
$$
\rig{(f^-_{1.1}(f^+_{1.1})'-(f^-_{1.1})'f^+_{1.1})}{T_3}-f^-_{1.1}(f^+_{1.1})'+f^-_{1.1})'f^+_{1.1}=-[2(f^+_{1.1}f^-_{1.1})'+({(f^-_{1.1})''\over f^-_{1.1}})' +
$$
$$
{1\over 2}{(f^-_{0.1})'f^-_{1.0}-(f^-_{1.0})'f^-_{0.1})\over f^-_{1.1}}{(f^-_{1.1})'\over f^-_{1.1}} +{1\over 2}{(f^-_{0.1})''f^-_{1.0}-(f^-_{1.0})''f^-_{0.1})\over f^-_{1.1}}+
$$
$$
{1\over 2}(f^+_{1.0}f^-_{1.0}+f^+_{0.1}f^-_{0.1})'+(f^+_{1.0}f^-_{1.0}+f^+_{0.1}f^-_{0.1}){(f^-_{1.1})'\over f^-_{1.1}} +{1\over 2}{f^-_{1.0}f^-_{0.1}\over f^-_{1.1}}{(f^-_{1.0}f^-_{0.1})'\over f^-_{1.1}}]
$$
$$
(\rig{f^-_{1.0}(f^+_{1.0})'-(f^-_{1.0})'f^+_{1.0}}{T_3})-f^-_{1.0}(f^+_{1.0})'+f^-_{1.0
})'f^+_{1.0}={f^-_{1.0}f^-_{0.1}\over f^-_{1.1}}[-2{(f^-_{1.0})''\over f^-_{1.0}}+{(f^-_{1.1})''\over f^-_{1.1}}+
$$
$$
{(f^-_{1.0})'f^-_{0.1}\over f^-_{1.1}}+2{(f^-_{1.0})'(f^-_{0.1})'\over f^-_{1.0}f^-_{0.1}}-{(f^-_{1.1})'\over f^-_{1.1}}({(f^-_{1.0})'\over f^-_{1.0}}+{(f^-_{0.1})'\over f^-_{0.1}})]+
$$
$$
2{(f^-_{1.1})'\over f^-_{1.1}}f^+_{0.1}f^-_{0.1}+(f^+_{0.1})'f^-_{0.1}-f^+_{0.1}(f^-_{0.1})'-f^-_{1.0}(f^+_{1.0})'+(f^-_{1.0})'f^+_{1.0}
$$
$$
(\rig{f^-_{0.1}f^-_{1.0}f^+_{1.1}-f^+_{0.1}f^+_{1.0}f^-_{1.1}}{T_3})-f^-_{0.1}f^-_{1.0}f^+_{1.1}+f^+_{0.1}f^+_{1.0}f^-_{1.1}=
$$
$$
{f^-_{1.0}f^-_{0.1}\over f^-_{1.1}}[{(f^-_{1.1})''\over f^-_{1.1}}+4{(f^-_{1.0})'(f^-_{0.1})'\over f^-_{1.0}f^-_{0.1}}-
$$
$$
2({(f^-_{0.1})'\over f^-_{0.1}}+{(f^-_{1.0})'\over f^-_{1.0}}){(f^-_{1.1})'\over f^-_{1.1}}-
{1\over 2}{(f^-_{0.1})'f^-_{1.0}-(f^-_{1.0})'f^-_{0.1})\over f^-_{1.1}}]+
$$
$$
2((f^-_{1.0})'f^+_{1.0}-(f^-_{0.1})'f^+_{0.1})+{(f^-_{1.1})'\over f^-_{1.1}}
(f^-_{0.1}f^+_{0.1}-f^-_{1.0}f^+_{1.0})
$$

\end{document}